\documentclass[superscriptaddress,showpacs,aps, prb, twocolumn]{revtex4-1}
\usepackage{times}
\usepackage{amsmath}
\usepackage{amssymb}
\usepackage{graphicx}
\usepackage{ulem}
\usepackage{pst-all}
\usepackage[colorlinks=ture,linkcolor=blue,citecolor=blue,urlcolor=blue]{hyperref}
\begin{document}
\title{ Neutral excitation and bulk gap of the Fractional Quantum Hall liquids in disk geometry} 
\author{Wu-Qing Yang}
\affiliation{Department of Physics, Chongqing University, Chongqing, P. R. China, 401331}
\author{Qi Li}
\affiliation{Department of Physics, South University of Science and Technology of China, Shenzhen, P. R. China, 518055}
\affiliation{School of Physics and Technology, Wuhan University, Wuhan, P. R. China, 430072}
\author{Lin-Peng Yang}
\author{Zi-Xiang Hu}
\email{zxhu@cqu.edu.cn}
\affiliation{Department of Physics, Chongqing University, Chongqing, P. R. China, 401331}
\date{\today} 

\begin{abstract}
For the numerical simulation of the fractional quantum Hall effects on a finite disk, the rotational symmetry is the only symmetry that were used in diagonalizing the Hamiltonian. In this work, we propose a method of using the weak translational symmetry for the center of mass of the many-body system.  With this approach, the bulk properties , such as the energy gap and the magneto-roton excitation are exactly consistent with that in the closed manifold like sphere and torus. As an application, we consider the FQH phase and its phase transition in the fast rotated dipolar fermions. We thus weapon the disk geometry versatility in analyzing the bulk properties beside the usual edge physics. 
\end{abstract}
\maketitle

\section{Introduction}
    The Fractional Quantum Hall Effect (FQHE), a topological quantum state of matter which was experimentally realized in two dimensional electron gas placed in a low temperature and strong magnetic field environment\cite{Tsui}, has attracted strong interest of theoretical and experimental physicists due to its embedded nature of the electronic topology and strong correlation.  Since the kinetic energy of electrons has been frozen by a strong magnetic field, the FQHE system is typically strongly correlated and cannot be treated by perturbation approach. The most powerful numerical tool for studying the FQHE is exact diagonalizing a microscopic Hamiltonian for small number of electrons or other advanced numerical methods, such as DMRG~\cite{JZDMRG, HuDMRG} or MPS~\cite{MPS}. Theoretically, the numerical calculation can be applied in different geometries for different purposes.  For the case of compact geometries without edge, such as putting electrons on the surface of a sphere or a torus, one always consider the bulk topological properties of the FQH states, i.e., the ground state topological degeneracy and magneto-roton excitation~\cite{GMP, boroton}. The open boundary systems, such as the cylinder and disk geometries,  are always aimed for the exploration of the edge physics~\cite{Wenedge}, such as the edge tunneling~\cite{LiJPCM, HuNJP}, quasiparticle interference~\cite{Willett0} and edge reconstruction~\cite{Wanedge}.  Thanks for the bulk-edge correspondence,  the topological properties of the FQH state can be unveiled from both the bulk and edge perspectives. 
    
   For the compact geometries, the proposal of the entanglement spectrum by Li and Haldane~\cite{HaldaneES} supplies a way of detecting the edge physics of the FQH droplets by artificially bi-partite the system. The entanglement spectrum is actually the eigenvalue spectrum of the reduced density matrix for subsystem after truncating the rest part of the system. It reflects the bulk topology via the counting of the so-called conformal edge states. However, the entanglement spectrum cannot give the quantitative properties of the edge excitation of the FQH droplet, such as the edge velocities or edge reconstruction.  In a realistic system, the existence of an edge is unavoidable. There are more parameters such as the strength of the background confinement and edge potential to tune the system. With these knobs, the nature of the FQH edge are extremely explored.   The edge excitation is gapless, which overwhelms the bulk excitation in the low-energy sector. The bulk excitation, such as the magneto-roton, is rarely discussed in disk geometry and therefore, the topological phase transition which accompanying gap closing in the bulk has been discussed incompletely. 
   
In this paper, we give a way for digging out the magneto-roton excitation of the FQH liquid in disk geometry and exactly determining the bulk energy gap by using the degrees of freedom in center of mass.  For the model Hamiltonian with $V_1$ interaction which gives the unambiguous Laughlin state and its low-lying excitations, we find the spectrum of magneto-roton excitation is exactly matched with that from the sphere geometry. The energy gap is found to be less sensitive to the finite size effect in our approach. As an example, we consider the dipolar interacted neutral atoms in a fast rotated trap~\cite{cooper08}. A phase transition from the FQH region to molecular phase as tilting the dipolar angle is characterized by the gap closing. In the anisotropic FQH state before gap closing, we observe multiple branches in the magneto-roton excitation.  The rest of this paper is organized as following:  In Sec. II, we review the model of the FQH system and the edge spectrum in orbital space on a disk. The center of mass diagonalization is introduced in Sec. III. We make a comparison of the energy gap with the spherical geometry and the magneto-roton excitation with torus geometry.  As an example of application, we consider the dipole-dipole interaction between the neutral atoms in a fast rotated trap in Sec. IV, in which the bulk gap closing manifests the phase transition and the multi-branch of the magneto-roton excitation  indicates the anisotropy of the FQH state.  Conclusions and discussions are arranged in the last section.

\section{Cyclotron motion in a magnetic field}
 For an electron in a magnetic field along $z$ direction, $B_z = \epsilon_{jk}\partial_j A_k$, the single particle Hamiltonian is
 \begin{eqnarray}
  \hat{H} = \frac{(\hat{p} - e\vec{A})^2}{2m} = \frac{\vec{\Pi}^2}{2m}.
 \end{eqnarray}

The coordinates of an electron in a magnetic field can be decomposed into the guiding center coordinates and cyclotron coordinates, namely $\vec{r} = \vec{R} + \tilde{R}$. The cyclotron coordinates are defined by the canonical momentum:
 \begin{eqnarray}
  \tilde{R}^a = -\frac{1}{eB} \epsilon_{ab} \vec{\Pi}_b.
 \end{eqnarray}
  It is easy to know
 \begin{eqnarray} \label{gccommu}
  \left[R_x, R_y\right] &=& -il_B^2, \nonumber \\
  \left[\tilde{R}_x, \tilde{R}_y\right] &=& il_B^2  \nonumber \\
  \left[R_i, \tilde{R}_j\right] &=& 0
 \end{eqnarray}
 
With these two types of coordinates, we can construct corresponding ladder operators
\begin{eqnarray} \label{aabb}
 a &\equiv&  \frac{1}{\sqrt{2}l_B} (\tilde{R}_y - i \tilde{R}_x) \nonumber \\
 b &\equiv&  \frac{1}{\sqrt{2}l_B} (R_x - i R_y) \nonumber \\
\end{eqnarray}
they obey the following commutation relations:
\begin{eqnarray}
 [a, a^\dagger] &=& [b, b^\dagger] = 1 \nonumber \\
 \left[a, b\right] &=& [a^\dagger, b]  = 0
\end{eqnarray}

\begin{figure}
 \includegraphics[width=8cm]{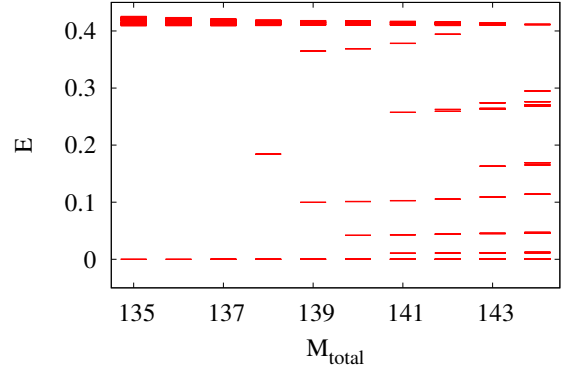}
 \caption{\label{v1spectrum}The spectrum of $V_1$ Hamiltonian for 10 electron in 30 orbitals in disk geometry. The unique zero energy state at $M_{total} = 135$ is the Laughlin state and the low-lying states in $M_{total} > 135$ are the edge excitation states.}
 \end{figure}

The Hamiltonian can be written in a diagonal form of one dimensional harmonic oscillator $H = (a^\dagger a + 1/2)\hbar \omega_c $ and the Hilbert space can be expanded by the above two ladder operators
\begin{eqnarray}
 |N, m \rangle = \frac{(a^\dagger)^N(b^\dagger)^m}{\sqrt{N!m!}}|0\rangle.
\end{eqnarray}
The index $N$ labels the landau level (LL) and $m$ labels the degenerated orbitals in each LL.
 After projecting into the lowest Landau level, only the orbital angular momentum quantum number $m$ remains and the system can be treated as a quasi-1D system for $N_e$ electrons filling $N_{orb}$ orbitals.  Because of the rotational symmetry along $z$-axis, the total angular momentum $M_{total} = \sum_i m_i$ is conserved and mostly used in the exact diagonalization. Fig.~\ref{v1spectrum} shows the energy spectrum with $V_1$ Hamiltonian in the language of the Haldane's pseudopotential~\cite{HaldanePP}  for $10$ electrons in $30$ orbitals. This is the model Hamiltonian for the celebrated Laughlin state~\cite{Laughlin} at filling factor $1/3$. The Laughlin state $\Psi_L = \prod_{i<j}(z_i - z_j)^3 e^{-\sum_i |z_i|^2/4l_B^2}$ has total angular momentum $M_{total} = M_L =3 N_e (N_e - 1)/2 = 135$ for $10$ electrons. The low-lying excited states that only exist in the subspace $M_{total} > M_L$ are the chiral edge states. Here we should notice that the edge excitations have zero energies in an infinite plane for the model Hamiltonian. The finite number of orbitals results in finite energies  for partially edge states while  $M_{total} > M_L + N_{orb} - 3 N_e + 2$.
 
 \section{Center of mass diagonalization}

 The information from the edge spectrum of the FQH state, especially for the realistic Coulomb interaction, sometimes does not completely specify the FQH state itself.  For example, we know the incompressible Laughlin phase at $1/3$ filling is accompanied by the root configuration ``100100...1001001"  which has edge counting ``$1, 1, 2, 3, 5 \cdots$" related to the chiral Luttinger liquid theory of the edge. However, some compressible charge density wave state, such as the largest anti-squeezed pattern ``$111\cdots000\cdots111$", could also be the global ground state and has exactly the same counting numbers because of the integer quantum Hall like edge near the edge.  Therefore, the edge spectrum is just one aspect of the topological order for the FQH liquids.   On the other hand, the topological quantum ground state is always defined by a bulk gap which is the origination of the incompressibility for the FQH liquids. The ground state phase transition is displayed by the gap closing.  Besides the gapless edge excitations on the boundary, the FQH liquids are embedded some bulk excitations. The most studied ones are the quasihole and quasielectron excitations which can be realized by a local potential~\cite{Hu08} mimic the STM or AFM tip in experiments. Relatively little progress has been made in understanding the neutral excitations since the seminal work of Girvin, MacDonald, and Platzman ~\cite{GMP} introduced the single-mode approximation (SMA) to describe the lowest excitation in terms of a neutral density wave, or the ``magnetoroton". For numerical simulation, the neutral excitation can be observed as an in-gap mode in the spectrum of a translational invariant torus geometry in both directions or rotational invariant sphere in the total angular momentum $L = 0$ subspace.   The bulk gap of the FQH liquid is defined as the gap between the roton minimum and the ground state. Thus the precise definition of the bulk gap, or finding out the roton minimum, is important to locate the phase transition of the ground state.   

In the energy spectrum in $M_{total}$ subspace on a disk, due to the gapless of the FQH edge excitation, we can not exactly locate the bulk gap closing except the changing of the global ground state due to edge reconstruction~\cite{Wanedge} in the lower energy sector.  For electrons on a finite disk, the translational symmetry is broken near the edge and the open boundary can be regarded as an infinite potential barrier. The Laughlin state is the most densest zero energy state for $N_e$ electrons in $3 N_e - 2$ orbitals for $V_1$ interaction, thus all the electrons prefer to form a droplet at the center of the disk. If the ground state is a homogeneous liquid state, we would expect that the center of mass (COM) locates at the center of disk, or the angular momentum for COM should be zero which is similar to the total angular momentum $L=0$ for the spherical geometry. The COM ladder operator for a many-body state is defined as the summation of all the operators in each guiding center orbitals.
\begin{figure}
\includegraphics[width=1.0\linewidth]{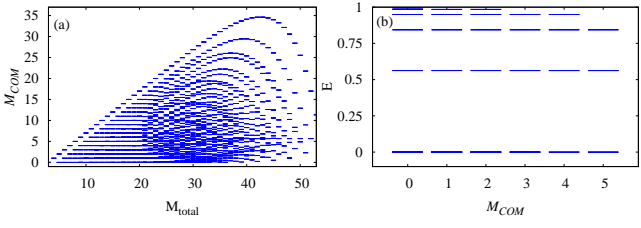}
\caption{\label{COMspectrum}The spectrum of the COM operator and energy spectrum for 3 electrons in 20 orbitals with $V_1$ interaction.}
\end{figure}
\begin{eqnarray}
 B = \frac{1}{N_e}\sum_i b_i
\end{eqnarray}
Therefore, the COM for a given state can be calculated as an expect value of its number operator $\hat{M} = B^\dagger B$:
\begin{eqnarray}
 M_{COM} = \langle \Psi | \hat{M} |\Psi\rangle.
\end{eqnarray}

\begin{figure}
\includegraphics[width=1.0\linewidth]{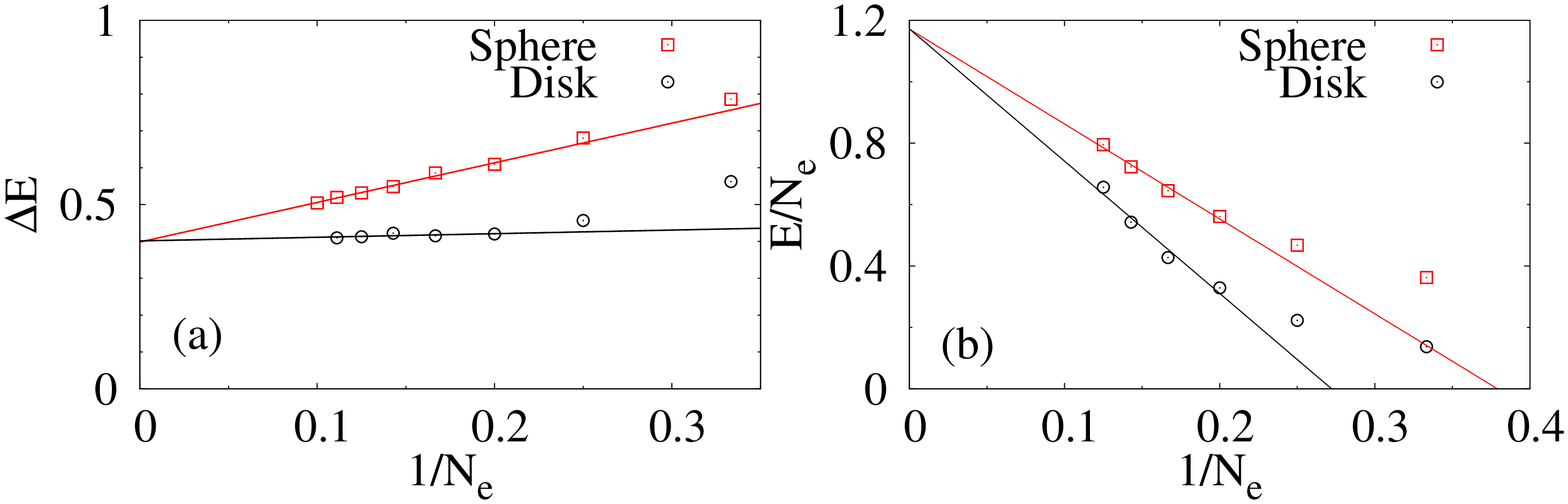}
\caption{\label{gap}The comparison of the band gap for the $V_1$ interaction (a) and ground state energy for Coulomb interaction (b) between disk and sphere geometries.}
\end{figure}
\begin{figure}
 \includegraphics[width=1.0\linewidth]{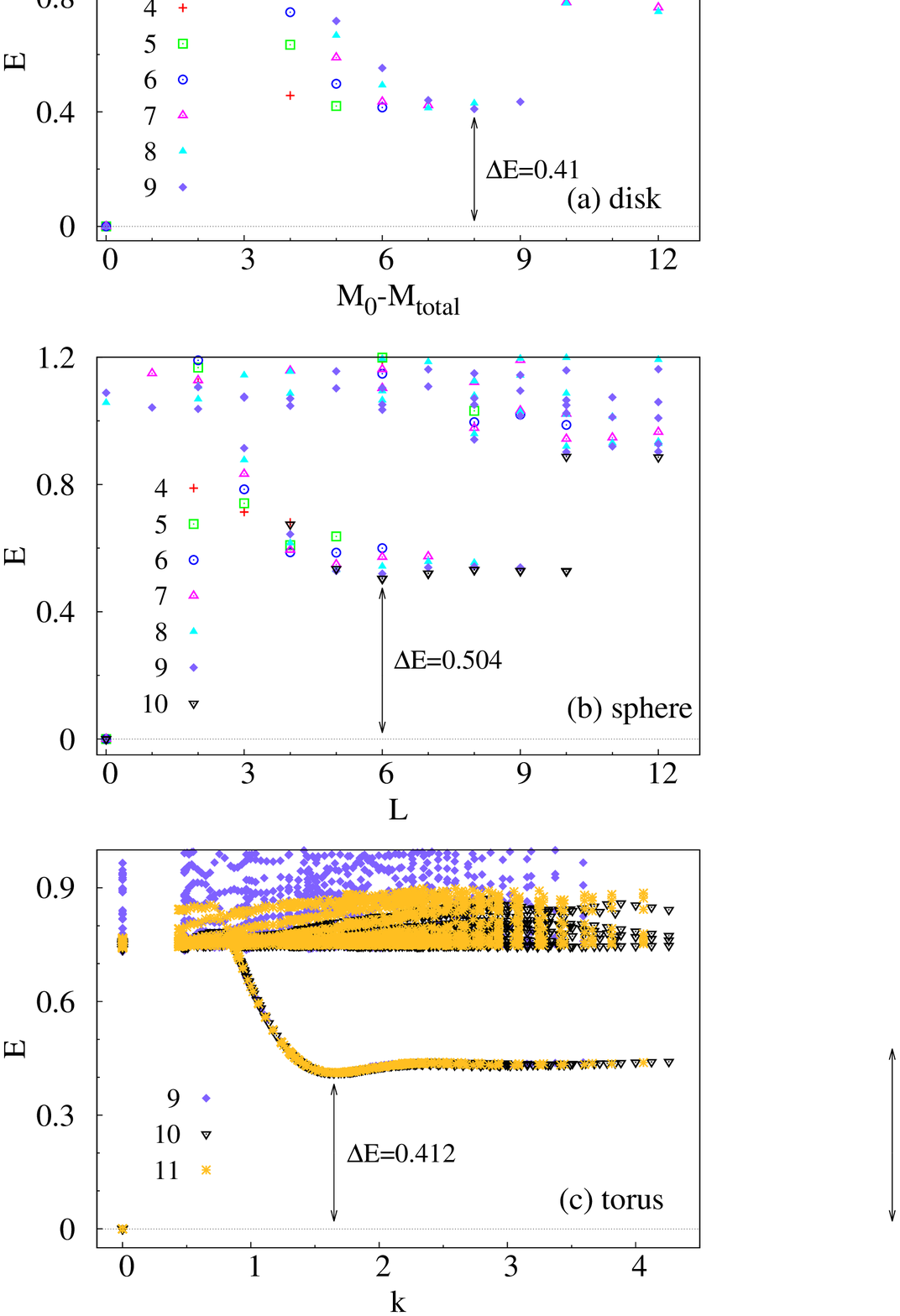}
 \caption{\label{roton}The magneto-roton spectrum on disk (a), sphere (b) and torus (c) for $V_1$ Hamiltonian .}
\end{figure}

For a infinite plane with translational invariance, the $[\hat{M}, H] =0$ and thus $M_{COM}$ is a good quantum number. We expect that the energy spectrum in different $M_{COM}$ subspaces should be degenerate. The presence of the edge for a finite disk will lift this degeneracy when the COM operator has contribution from edge.  As shown in Fig.~\ref{COMspectrum}(a), we diagonalize the  $\hat{M}$ operator in a system with 20 guiding center orbitals with the quantum number $M_{total}$. The eigenvalues of the $\hat{M}$ are well quantized to integers while $M_{total} < N_{orb}$. If we make a truncation to keep all the well quantized states in the spectrum of $\hat{M}$, the electron Hamiltonian can be approximately diagonalized in each truncated $\hat{M}$ subspace. The results for 3 electrons in 20 orbitals are shown in Fig.~\ref{COMspectrum}(b).  It shows that the spectrum is highly degenerated for different $\hat{M}$ subspaces. Some high energy excited states gradually vanish as increasing $M_{COM}$. Therefore, from the experience of the spherical geometry, we speculate that the spectrum in $M_{COM} = 0$ subspace contains all the information in the bulk. First of all, the ground state energy is exactly zero for $V_1$ interaction as the way we would expect. Second, the first excited state defines the bulk gap. As shown in Fig.~\ref{gap}, we compare the bulk gap for the $V_1$ interaction and the ground state energy per particle for Coulomb interaction with that on sphere for several system sizes. It is found that the thermodynamic limit extrapolated values are consistent for different geometries. However, it is obvious that both the energy gap ($V_1$) and single particle energy (Coulomb) in disk geometry are always smaller than that on sphere. Moreover, the gap has much fewer finite size effects in disk geometry.  It reaches the saturation value even at a very small size for 5 electrons with $V_1$ interaction. At last, except the bulk gap, we can also extract the neutral magneto-roton excitation from the energy spectrum in $M_{COM} = 0$ subspace. Because the rotational invariance is always there, we calculate the expect value of $M_{total}$ for each states in $M_{COM} = 0$ subspace and replot the spectrum as a function of $M_{total}$. The results are shown in Fig.~\ref{roton}(a).  While we plot the spectrum by $M_0 - M_{total}$ where $M_0$ is the total angular momentum for ground state, i.e., $M_0 = 3 N_e (N_e - 1)/2$, the magneto-roton comes out.  Thus the magneto-roton excitation actually hidden in the $M_{total} < M_0$ subspace in Fig.~\ref{v1spectrum}.  As a comparison, we plot the energy spectrum for the same $V_1$ Hamiltonian in sphere (with the total angular momentum $L$) and torus geometry (with the total translational momentum $k$) as shown in Fig.~\ref{roton}(b) and (c) respectively. The two closed manifolds clearly show the magneto-roton excitation branch in the spectrum. As we can see, the curves for the neutral excitations are almost the same although the system size we can do in the disk geometry is smaller. Because the full diagonalization of the COM matrix to construct the $M_{COM}=0$ basis limits the system size.

Therefore, as we can see, all the bulk properties of the FQH liquid could be extracted from a finite disk. All we just need is to diagonalize the electron Hamiltonian in the $M_{COM} = 0$ subspace. The ground state, the gap and the neutral excitation in the bulk can be obtained. Because of no curvature, the gap in disk geometry has much fewer finite size effects.

\section{Phase transition of the dipolar atoms in FQH regime}
\begin{figure}
\includegraphics[width=6cm]{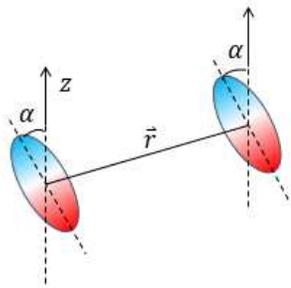}
\caption{\label{Modeldipole} The model for the dipolar Fermions.  $\alpha$ is the angle between the
direction of the dipole moment and $z$ axis.  The system is placed in a confining potential $\gamma m$ in the $m$'th orbital. }
\end{figure}

In the Bose-Einstein condensation of $^{52}\text{Cr}$ atom~\cite{PhysRevLett.94.160401} or the degenerate quantum gas of $^{40}\text{K}^{87}\text{Rb}$~\cite{PhysRevLett.78.586}, the interaction can be described as dipole-dipole interaction with the $s$-wave  collisional interaction vanishing for spin polarized fermions.  We assume all the dipoles are polarized in the same direction and without loss of generality, say $x-z$ plane. The polarized dipole interaction (in unit of $d^2/(4\pi\epsilon_0 l^3)$, where $d$ is the dipole moment of the neutral atom and $\epsilon_0$ is the vacuum permittivity) is
\begin{equation} \label{vddr}
 V_{dd}(\bold{r},\alpha) = \frac{1-3(\bold{\hat{d}}\cdot\bold{\hat{r}})^2}{r^3} = \frac{r^2 - 3(z\cos\alpha + x\sin\alpha)^2}{r^5}
\end{equation}
with $\bold{\hat{d}}$ being the direction of the polarized dipole (parametrized by the angle $\alpha$) as sketched in Fig. ~\ref{Modeldipole}.   When the dipole fermions are placed in a fast rotated trap and the rotating frequency is approaching the trap frequency along $z$ axis, the Coriolis force experienced by the atoms results in an effective magnetic field. Then the FQH states are expected in the fast rotating limit.~\cite{cooper08, RMP81647,PhysRevLett.94.070404,oster07, baranov08}   By using our recently developed generalized pseudopotential description of the anisotropic FQH state~\cite{bohu17, YangCPB}, we studied~\cite{HuPRB2018} the anisotropic FQH state and its quantum phase transition as tilting the dipolar angle $\alpha$. In torus geometry,  it was shown that
the gap of the Laughlin-like state at $1/3$ filling was closed at $\alpha \simeq 53^\circ$ when the thickness was set to $q = 0.01$.  Previous studies in disk geometry~\cite{Qiu2011} could not get this critical value because of the edge effects.

\begin{figure}[htb]
\includegraphics[width=1.0\linewidth]{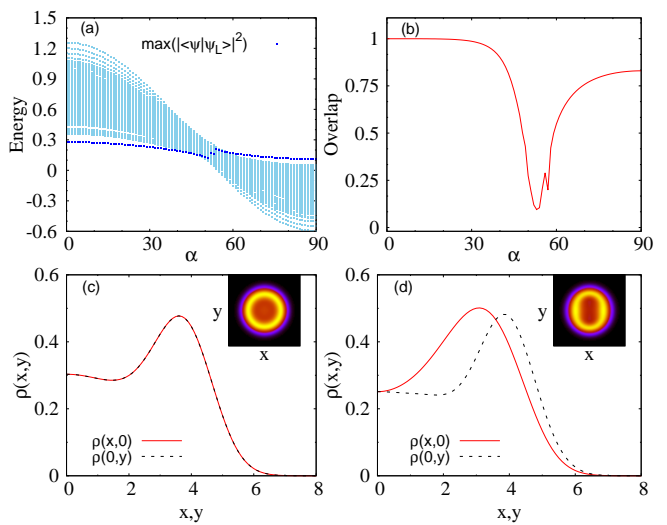}
\caption{\label{dipolegap} (a)The energy spectrum for 5 dipolar fermions in 15 ortitals with the confinement potential $\gamma=0.005 $ as varying the tilting angle $\alpha$. The states labelled by dark blue points have the maximum overlap with Laughlin state. (b)The largest overlap along the dark blue points. The particle density and their profiles along $x/y$ directions for the ground state at $\alpha=0^{\circ} $ (c) and highest energy state at $\alpha=90^{\circ} $ (d).}
\label{aniso-cross}
 \end{figure} 

In this section, we want to provide complementary numerical results about this phase transition in the disk geometry. In this case, we have one additional parameter, the confining potential strength $\gamma$ which characterizes the relative strength of confining potential with respect to interaction.  In our calculation, we always choose the value of $\gamma$ to maximize the energy gap.  The pseudopotential analyze~\cite{HuPRB2018} tells us that the dipole-dipole interaction in the lowest Landau level can be described by a model Hamiltonian $V_1 + \lambda V_{1,2}$. Therefore, as the $V_1$ model in Sec. III, we expect the finite size effect is small in the dipole-dipole interaction and the results for small system should be good enough to predict the physics in the thermodynamical limit. On the other hand, the rotational invariance is broken while the dipole moment has component in the plane, we thus need to include the orbitals as many as possible in order to get convergence of the energy.

 \begin{figure}
  \includegraphics[width=1.0\linewidth]{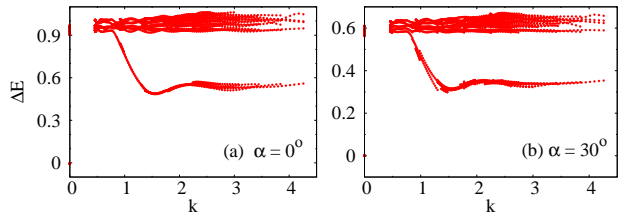}
 \caption{\label{rotondipole}The energy spectrum on torus for 10 dipolar fermions at $1/3$ filling in LLL at $\alpha=0^{\circ} $ (a) and $\alpha=30^{\circ} $ (b).}
 \end{figure} 
 
In Fig.~\ref{dipolegap}(a), we plot the energy spectrum in the $M_{COM} = 0$ subspace for 5 particles in 15 orbitals with confinement $\gamma = 0.005$. It shows clearly that the bulk gap between the ground state and the first excited state gradually decreases as increasing the 
angle $\alpha$.  The gap closes at $\alpha \simeq 53^\circ$ while $q = 0.01$ which is exactly the same as that from the calculation in torus geometry~\cite{HuPRB2018}.   To trace the ground state, we calculate the largest wave function overlap between the Laughlin state and the eigenstates in each subspace.   Predictably, the ground state before gap closing has the largest overlap with the model wave function. Here we just consider the isotropic Laughlin state as a reference, it is shown in Fig.~\ref{dipolegap}(b) that the overlap is almost one even though the tilted angle $\alpha$ reaches to $30^\circ$ which demonstrates that the ground state is still isotropic. If we continue to trace the highest overlap state in the spectrum, as outlined in blue, it looks like the energy of the Laughlin-like state flows up and become the highest energy state finally. When $\alpha = 90^\circ$, we find the highest state has overlap $\sim 75\%$ with Laughlin state.  In Fig.~\ref{dipolegap}(c) and (d), we plot the particle density and their profiles along $x$/$y$ directions for the ground state at $\alpha = 0^\circ$ and highest state at  $\alpha = 90^\circ$.   When $\alpha = 0^\circ$, the droplet is isotropic as expected.  For $\alpha = 90^\circ$, the shape of the density  is elliptical and has different profiles along $x$ and $y$ directions.  We should note that the reference state in the overlap calculation is the isotropic Laughlin state which does not contain the geometric metric~\cite{HaldaneGeometry}.  This anisotropic Laughlin-like state should have larger overlap with the anisotropic model wave function~\cite{Qiuprb12}.   

In spite of the COM diagonalization can provide the bulk gap and the outline of the neutral magneto-roton excitation, since the angular momentum is discreted, we do not find a way to get the continue spectrum for the roton excitation as that in torus (The reason that one can get the continue spectrum on torus is the angle between two primitive lattice vector can smoothly be tuned). Here, as an attachment, we plot the magneto-roton spectrum for the dipolar fermions at $1/3$ filling in the LLL in torus geometry. As shown in Fig.~\ref{rotondipole},  when $\alpha = 0^\circ$, the spectrum looks similar to the $V_1$ model in Fig.~\ref{roton}(c) except that the long range interaction makes the points scattered for large $k$ part.  When $\alpha = 30^\circ$ at which the ground state still looks isotropic since it has almost $100\%$ overlap with the isotropic model wave function as shown in Fig.~\ref{dipolegap}(b). However, the neutral excitation in the bulk as shown in Fig.~\ref{rotondipole}(b) has some multi-branch structures. This result is similar to the case while considering the effects of anisotropic effective mass on the FQH state~\cite{BoYangPhysRevB.85.165318}.  It is reasonable since we know both the anisotropic effective mass and the anisotropic interaction could be equally treated as an intrinsic metric in the geometric description of the FQH states~\cite{HaldaneGeometry}.
  
\section{Summaries and conclusions}
In conclusion, we find a numerical method to extract the bulk properties of the FQH liquids, such as the energy gap and the neutral magneto-roton excitation in the disk geometry which is mostly used to study the edge physics in the guiding center orbitals.  The main point in our approach is constructing the electron Hamiltonian in the subspace of the COM angular momentum. Although the COM angular momentum is conserved in an infinite plane,  we find the diagonalization of the interaction Hamiltonian in a truncated $M_{COM}=0$ subspace provides all the information in the bulk.  With comparing the results in closed geometries, such as sphere and torus which naturally provides the bulk properties, we find the extrapolated energy gap in thermodynamic limit is consistent both for the model Hamiltonian and the Coulomb interaction. More importantly, because of no curvature in the plane, the finite size effect is smaller than that on sphere.  The magneto-roton excitation comes out while all the eigenenergies in the $M_{COM}=0$ subspace are plotted as the guiding center angular momentum $M_{total}$ which looks very similar to the spectrum on the sphere with considering the total angular momentum $L$.  As an application, we consider the fast rotated dipolar fermions in quantum Hall regime.  The critical point of the gap closing for the phase transition induced by tilting the dipole angle is consistent to the previous study on torus.  We also find that the Laughlin state in the COM spectrum flows to be the highest energy state when all the dipoles are parallel to plane although it comes anisotropic.  The multi-branch structure of the magneto-roton excitation in the anisotropic FQH state is  also consistent with the previous study of the anisotropic effective mass. The weakness of this method is the system size has an upper limit  due to the full diagonalization of the COM matrix.

\begin{acknowledgements}
This work is supported by National Natural Science Foundation of China Grants No.11674041, 91630205, 11847301, Chongqing Research
Program of Basic Research and Frontier Technology Grant No. cstc2017jcyjAX0084 and FRF for the Central Universities No. 2019CDJDWL0005. 
Q. Li thanks for support of the National Natural Science Foundation of China Grant No.11474144.
 \end{acknowledgements}

\end{document}